\documentclass[12pt,fullpage,fleqn,psfig]{article}

\usepackage{latexsym} 
\usepackage{epsfig}
\large
 \newcommand{\beq}{\begin{equation}} 
\newcommand{\eeq}{\end{equation}} \newcommand{\ba}{\begin{array}{ccc}}

\newcommand{\ea}{\end{array}}
  
\newcommand{\be}{\begin{eqnarray}}
\newcommand{\ee}{\end{eqnarray}}

\begin{document}
\newcommand{\mat}{\left ( \begin{array}{cc}} \newcommand{\emat}{\end{array} \right )}
                                                        
\newcommand{\matt}{\left ( \begin{array}{ccc}}
    \newcommand{\ematt}{\end{array} \right )} \newcommand{\matf}{\left ( \begin{array}{cccc}}
    \newcommand{\ematf}{\end{array} \right )} \newcommand{\vect}{\left ( \begin{array}{c}}
    \newcommand{\evect}{\end{array} \right )} \newcommand{\Tr} {\rm
  Tr}  \newcommand{\cotanh}{{\rm cotanh}} \def\beqn{\begin{eqnarray}} 
 \def\eeqn{\end{eqnarray}} \def\la{\lambda} \def\ga{\gamma}
\def\om{\omega}   \def\al{\alpha} \def\d{\partial} \def\Tr{ {\rm Tr} }

\thispagestyle{empty}
\parskip=4mm

\begin{flushright}
\hfill{SUNY-NTG-01/42}
\end{flushright}

\vspace{1cm}
\begin{center}
{\Large\bf Kaon Condensation and  Goldstone's Theorem} 

\vspace{8mm}

T. Sch\"afer$^{1,5}$, D.T. Son$^{2,5}$, M.A. Stephanov$^{3,5}$, 
D. Toublan$^4$ and J.J.M. Verbaarschot$^1$

{\small 
$^1$Department of Physics and Astronomy, SUNY, Stony Brook, NY
11794}\\  {\small $^2$Physics  Department, Columbia University, New
York, NY 10027}\\  {\small $^3$Department of Physics, University of
Illinois, Chicago, IL 60607}\\ {\small $^4$Physics Department,
University of Illinois at Urbana-Champaign, Urbana, IL 61801}\\
{\small $^5$RIKEN-BNL Research Center, Brookhaven National Laboratory,
Upton, NY 11973}
\vskip 1.5cm

{\bf Abstract}

\end{center}

We consider QCD at a nonzero chemical potential for strangeness. At a
critical value of the chemical potential equal to the kaon mass, kaon
condensation occurs through a continuous phase transition. We show
that in the limit of exact isospin symmetry a Goldstone boson with the
dispersion relation $E \sim p^2$ appears in the kaon condensed phase. 
At the same time, the number of the Goldstone bosons is less than the 
number of broken generators. Both phenomena are familiar
in non-relativistic systems.
We interpret our  results in terms of a Goldstone boson counting rule
found previously by Nielsen and Chadha. We also formulate a criterion
sufficient for the equality between the number of Goldstone bosons and
the number of broken generators.

\vskip 0.5cm
\noindent
{\it PACS:} 11.30.Rd, 12.39.Fe, 12.38.Lg, 71.30.+h
\\  \noindent
{\it Keywords:} QCD partition function; Kaon Condensation; Finite
Isospin Density;  Goldstone's Theorem; Quadratic Dispersion Relations;
Low-energy  effective theory; Chiral Perturbation Theory

\newpage
\section{Introduction}
Because of the spontaneous breaking of chiral symmetry, QCD at low
energy reduces to a theory of weakly interacting Goldstone
bosons. In the  chiral limit the Lagrangian of this theory is
completely determined  by chiral symmetry and Lorentz invariance. Its
predictions  for the low-energy phenomenology of QCD have been very
successful \cite{GaL}.   
One might be tempted
to extend this theory to describe QCD at nonzero chemical potentials.  
For a baryon
chemical potential, the extension is rather trivial since the
Goldstone bosons do  not carry baryon charge.  Thus a small baryon
chemical  potential does not have any effect on the chiral
dynamics. The situation is more interesting for an isospin 
chemical potential. In this case a transition to a pion condensed phase
takes place at a chemical potential equal to 
the pion mass. For light quarks this phase transition takes place
within the domain  of validity of the chiral Lagrangian and is
completely described by means of a low energy effective theory   
\cite{Misha-Son} which also describes 
phase quenched QCD \cite{DT-phase}. Similarly, the low energy limit of
QCD at at nonzero  strange chemical potential, including its rich
phase diagram at zero temperature can be   
completely described in terms of a chiral Lagrangian \cite{KT}. 

In this letter we analyze the general properties of the spectrum of
Goldstone bosons in a symmetry-broken phase induced by a chemical
potential.  In the process, we will encounter a rather unusual
manifestation of Goldstone's theorem \cite{Goldstone,GSW},  
in which the number of Goldstone
bosons is less than the number of broken generators.  A comment on the
subtleties of Goldstone's theorem is thus in order. 

According to the most frequently encountered version of Goldstone's theorem, 
the number of Goldstone modes is equal to the 
number of independent broken symmetry generators \cite{Kibble}.  It is
worth noting, however, that the original formulation of Goldstone's
theorem is much weaker: it states that in the presence of broken
symmetries, there exists {\em at least one} massless mode.  In
relativistically covariant theories the number of Goldstone bosons is
{\em always} equal to the number of broken generators (we will give
our version of the proof later in the paper). There are, however,
well-known examples of nonrelativistic theories where the number 
of Goldstone modes is not equal to the number broken  
generators \cite{Sachdev,magnetic-leutwyler, Hofmann, foundations}.  
The simplest case is the Heisenberg ferromagnet, which has
only one magnon despite the fact that the symmetry breaking pattern is
$O(3)\to O(2)$.  [This is in contrast to the antiferromagnet where the
symmetry breaking pattern is the same but there are two magnons].  

  Goldstone's theorem has been refined in a little known article by
Nielsen and Chadha \cite{holger-counts}. They distinguish two types of
Goldstone bosons: those with an energy proportional to an
even power of the momentum and those with a dispersion relation that 
is an odd power of the
momentum. They formulated a theorem, which states that one has to
count each Goldstone mode of the first type {\em twice}. More
precisely, the sum of twice the number of Goldstone modes of the first
type and the number of Goldstone modes of the second type is at least
equal to the number of independent broken symmetry generators.   
For relativistic invariant systems the dispersion relation
for massless states
is necessarily linear and the total number of Goldstone bosons is 
equal to the number of broken symmetry generators. 
This theorem explains why one (instead of two) Goldstone boson appears in the Heisenberg 
ferromagnet: the dispersion relation for this boson
is quadratic in the momentum ($E\sim p^2$).
The so-called canted phase of
ferromagnets carry one Goldstone boson of each type \cite{Sachdev},
and require a complete breaking of $O(3)$.

Quantum field theory at nonzero chemical potential  is not Lorentz 
invariant.  This fact
opens up the possibility of having both Goldstone modes with 
a linear dispersion relation and a quadratic dispersion relation, and
the character of the Goldstone modes could change across a phase
transition. The number of Goldstone bosons and broken generators do
not need to coincide. In this letter we illustrate this possibility
for the example of QCD at a nonzero strangeness chemical potential.

In section 2 we
study  a simple field-theoretical model
that describes the interaction of the strange Goldstone
bosons in the neighborhood of the kaon-condensation point.
This model has been recently used to describe kaon condensation
\cite{kaon-condensation,Schaefer,BedaqueSchaefer,KaplanReddy}
in the
color-flavor locked phase of QCD \cite{CFL} at very high densities.
 Goldstone's theorem will be discussed in section 3.
In section 4 we analyze the low-energy limit of QCD at nonzero strangeness
chemical potential which was constructed in \cite{KT}.
Concluding remarks are made in section 5.

\section{Spectrum of Goldstone modes in a simple model}

In this section we consider the
simple (Euclidean) Lagrangian,
\be
  {\cal L} = (\partial_0 + \mu)\phi^\dagger
  (\partial_0 - \mu)\phi + \partial_i\phi^\dagger\partial_i\phi
  + M^2 \phi^\dagger \phi + \lambda 
  (\phi^\dagger \phi)^2,
  \label{Lagr}
\ee
where $\phi$ is a complex scalar doublet,
\be
  \phi = \vect \phi_1 \\ \phi_2 \evect.
\ee
The Lagrangian (\ref{Lagr}) possesses a $U(2)=SU(2)\times U(1)$
symmetry, and $\mu$ is the chemical potential with respect to the
$U(1)$ charge.  The $U(2)$ symmetry is essentially that of the Higgs
sector of the standard model.  The motivation for considering Eq.\ 
(\ref{Lagr}) at finite $\mu$ is two-fold. First,  from the physics of
kaon condensation in the color-flavor locked phase of  
QCD at high baryon density.  
In this phase, the lightest modes are the four charged and neutral
kaons, which are degenerate if up and down quarks have the same mass.
These kaons are described by the fields $\phi_1$ and $\phi_2$, while
other mesons are neglected.  The role of $\mu$ can be played by the
strangeness chemical potential or effectively by a  
strange quark mass \cite{BedaqueSchaefer}.  
Though somewhat simplified, Eq.\ (\ref{Lagr})
captures the essential physics of kaon  
condensation. Second, near the phase transition point, the low-energy
limit of QCD at nonzero strangeness   
chemical potential given by the effective theory constructed in
\cite{KT} can be reduced to the Lagrangian one involving only the
strange degrees of freedom. The reason is that, at the phase
transition point, only strange degrees of freedom become massless.

The chemical potential contributes $-\mu^2\phi^\dagger\phi$ to the
potential energy.  Here we consider only the case when $M^2>0$ so the
symmetry is unbroken at zero chemical potential.  Turning on $\mu$, we
start to favor quanta of, say, positive $U(1)$ charge
(``strangeness''), but if $|\mu| < \mu_c \equiv M$ we expect no change
since the chemical potential is not sufficient to excite any
quanta. Only when $\mu > \mu_c$ do we expect any condensation.  This
is exactly the regime where the potential energy has a nontrivial
minimum. The vacuum can be chosen to have 
\be  
 \phi = {1\over\sqrt2} \vect 0  
  \\ v \evect, \qquad {\rm with} \quad v^2 = \frac{\mu^2 -M^2}\lambda.
  \label{choice}
\ee 
The symmetry is thus broken from $U(2) \to U(1)$, resulting in three broken generators.

The particle spectrum is obtained by
expanding about the minimum of the potential. In the normal phase
($\mu<\mu_c$), with $\langle\phi \rangle = 0$, we find four modes, two
with strangeness $S=1$ and two with $S=-1$, and the dispersion
relations are given by 
\be  
  (E+S\mu)^2 = p^2 + M^2. 
\ee
At the transition point $\mu=M$, two modes become gapless 
(those with $S=+1$). By continuity, these modes become the Goldstone 
bosons in the broken phase, therefore there are only two Goldstone modes. 
As we will show, this is possible because one of the Goldstone bosons has 
a quadratic dispersion relation.

 To find particle dispersion in the broken phase, we expand the fields about 
the minimum as follows
\be
\phi = {1\over\sqrt2} e^{i\pi_k \tau_k/v} \vect 0 \\ v+\varphi
\evect. 
\ee 
The quadratic part of the Lagrangian has the form   
\be 
{\cal L} &=& \frac 12 \partial_\nu \varphi \partial_\nu \varphi 
+\frac {1}2 \partial_\nu \pi_k \partial_\nu \pi_k
-i\mu (\varphi \partial_0 \pi_3 - \pi_3 \partial_0  \varphi)
-i\mu (\pi_1 \partial_0 \pi_2 -\pi_2 \partial_0 \pi_1) +
v^2\lambda \varphi^2.
\nonumber\\
  \label{Lagr2}
\ee
The particle spectrum and dispersion relations thus follow from
diagonalizing the $2\times 2$ matrix that couples the $\varphi$ and
the $\pi_3$ modes,  
\be D_1= \mat p^2 -E^2 +2 v^2 \lambda &
                       2i \mu E \\ 
             -2i\mu  E  &p^2 - E^2 \emat.
\ee
and the $2\times 2$ matrix that couples the $\pi_1$ and the $\pi_2$ modes, 
\be
D_2 =\mat p^2 -E^2  & 2 i \mu    E \\
        -2i \mu  E & p^2 - E^2   \emat.
\ee 

First let us consider the sector of $\varphi$ and $\pi_3$.
The dispersion relations are obtained from
\be
\det D_1 = (p^2-E^2 -\lambda_1)(p^2-E^2-\lambda_2)=0. 
\ee
The eigenvalues $\lambda_1$ and $\lambda_2$ satisfy   
$\lambda_1 \lambda_2 = -4\mu^2 E^2$.
With $\lambda_1=-2 v^2 \lambda + O(E)$ we thus have that $\lambda_2 =
2\mu^2 E^2/ \lambda v^2 + O(E^3)$. In this case with only one broken
generator we thus find a quadratic $E$-dependence of $\lambda_2$
resulting in a linear dispersion relation. The small momentum
expansion of the dispersion relation is given by   
\be
  E^2= \frac{\mu^2-M^2}{3 \mu^2- M^2}\, p^2
             + {\cal O}(p^4)\,
              . \label{displ} \ee 
The other mode remains massive for $p \to 0$ with 
dispersion relation 
\be E^2 = 6\mu^2 -2M^2
              + {\cal O}( p^2)\, . 
\ee
Notice that the coefficient of $p^2$ in (\ref{displ})
vanishes at  $\mu=M$. As we will see below, this is a necessary
consequence of the continuity of the dispersion relation
across the phase transition point. 

\begin{center}
\begin{figure}[ht!]
\vspace{0.5cm}
\hspace*{1cm}
\epsfig{file=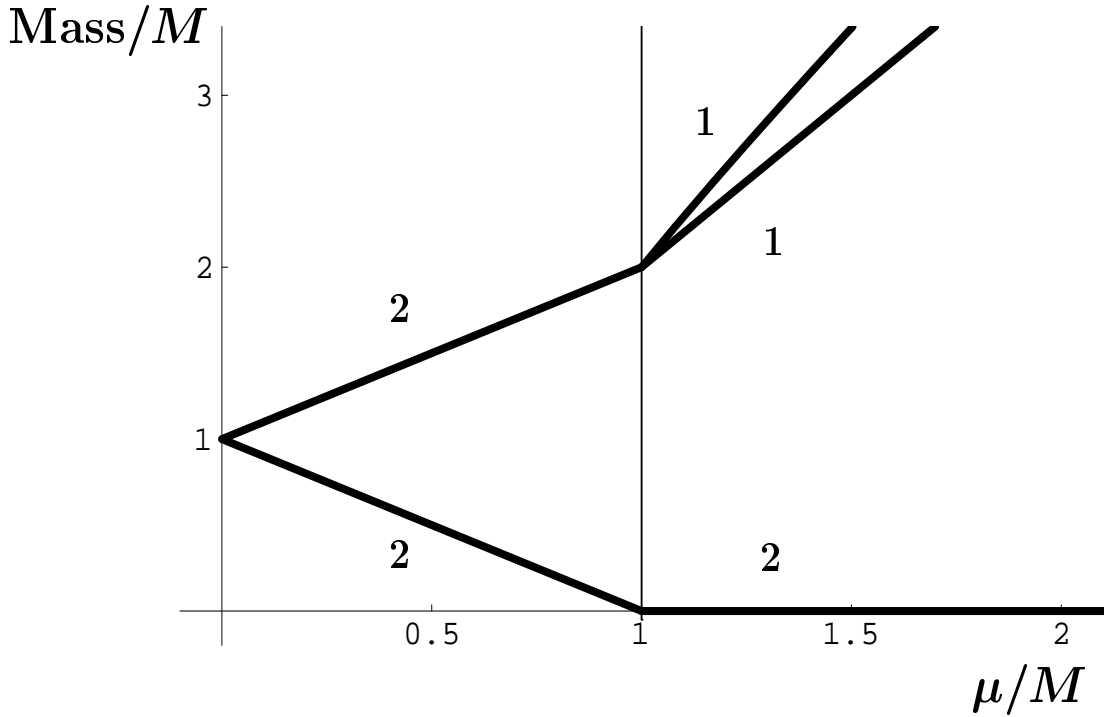, width=14.6cm, height=9.5cm} 
\caption[]{\small 
Spectrum of the excitations of (\ref{Lagr}) as a function
of $\mu$. The critical chemical potential is at $\mu=M$, and it is
depicted by a vertical line. The
degeneracies of each of the branches is given by the number
along the curves.} \label{fig:masses}
\end{figure} 
\end{center}

In the sector of $\pi_1$, $\pi_2$ the curvature matrix ($D_2$ at
$E=p=0$) has two zero
eigenvalues. However, only one of the pole masses vanish. This follows
immediately from the dispersion relations   
\be
E= 2\mu  + {\cal O}(p^2), \qquad E = \frac {p^2}{2\mu
  }+{\cal O}(p^4). \ee which are obtained from \be \det D_2 = (p^2-E^2
-\lambda_1)(p^2-E^2-\lambda_2) =0,  
\label{dispq}
\ee 
with 
\be
\lambda_1 = 2\mu  E, \qquad \lambda_2 = -2\mu  E. 
\ee 
We observe that in our theory the total number of Goldstone modes is 
consistent with the Nielsen-Chadha theorem \cite{holger-counts}.  The 
behavior of the gaps of different modes in our theory as a function 
of $\mu$ is illustrated in Fig.\ \ref{fig:masses}.

\section{Goldstone's theorem}

One might still be puzzled by the fact that both $\pi_1$ and $\pi_2$
in (\ref{Lagr2}) correspond to two flat directions of the potential
energy, but describe only one Goldstone mode.  One way to understand
this fact is to consider the limit of very low frequencies, much lower
than $\mu$. In this case the terms proportional to
$(\partial_0\pi_1)^2$ and $(\partial_0\pi_2)^2$ in Eq.\ (\ref{Lagr2})
are negligible compared to the
$\pi_1\partial_0\pi_2-\pi_2\partial_0\pi_1$.   
Once these terms are ignored one immediately sees that $\pi_1$ and
$\pi_2$ are two {\em canonically conjugate} variables, and hence
corresponds to a Goldstone field and its time derivative, but not two
different Goldstone fields. 

As mentioned above, Lorentz invariant theories can have only Goldstone
modes with $\omega=p$, i.e., with a linear dispersion relation.  The
inverse is, however,   
not true: there are many 
theories which are not relativistically invariant but have only
Goldstone modes with linear dispersion relations.  The most familiar
example is the antiferromagnet; there are also other examples
\cite{KSTVZ}. Since this class of theories seems to be rather wide,  
a more refined criterion to count the
Goldstone bosons beside Lorentz invariance could be useful. 
To distinguish the different possibilities we have formulated the following theorem: 

\noindent{THEOREM}.
If $Q_i$, $i=1,\ldots,n$ is the full set of broken generators, and if
$\langle0|[Q_i,Q_j]|0\rangle=0$ for any pair $(i,j)$, then the number
of Goldstone bosons is equal to $n$, i.e., the number of broken
generators.  

Indeed, assume that there are less than $n$ Goldstone bosons.  Since
the zero-momentum Goldstone bosons are obtained by acting the broken
generators on the ground state, there exist a linear combination of
$Q_i$ which annihilates the ground state, \begin{equation}  
  \sum_i a_i Q_i |0\rangle = 0.
  \label{annihilate}
\end{equation}
Equation (\ref{annihilate}) does not necessarily contradict the
assumption that there are $n$ broken generators, since the
coefficients $a_i$ are, in general, complex, so $\sum_i a_i Q_i$ may
be non-Hermitian and hence does not need to be a generator of the symmetry
algebra.  However, the following objects are generators,
\begin{eqnarray} 
  Q_a = \sum_i {\rm Re}a_i\, Q_i, \quad {\rm and }\quad
  Q_b = \sum_i {\rm Im}a_i\, Q_i,
\end{eqnarray}
and hence $Q_a|0\rangle\neq0$ and $Q_b|0\rangle\neq0$.
We will show that $\langle0|[Q_a, Q_b]|0\rangle\neq0$.  
Indeed, Eq.~(\ref{annihilate}) reads $(Q_a+iQ_b)|0\rangle=0$, 
hence we can write 
\begin{eqnarray} 
  Q_a |0\rangle = |b\rangle, \qquad
  Q_b |0\rangle = -i|b\rangle,
\end{eqnarray}
which is the definition of the state $|b\rangle$.  It is now trivial 
to see that 
\begin{equation} 
  \langle0|[Q_a,Q_b]|0\rangle = -2i\langle b|b\rangle \neq 0,
\end{equation} 
which contradicts the original assumption.  QED.  

Recalling that $[Q_i, Q_j]$ is a linear combination of the symmetry
generators (in general, both broken and unbroken), one can immediately
conclude that if the densities of all conserved charges are equal
zero, the number of Goldstone bosons is equal to the number of broken
generators.  To have a mismatch between the number of Goldstone bosons
and the broken generators, it is necessary (but not sufficient) to
have  
finite charge density.  For example, in ferromagnets the density of
the total spin is nonzero, so it is possible to have only one magnon
for two broken generators, while in antiferromagnets all densities
vanish, and hence there should be two magnons. In particular, if the
ground state $|0\rangle$ is Lorentz-invariant, all charge densities
should vanish (since they are zeroth components of the corresponding
four-currents), and there should be no mismatch.  

In our example, there are three broken generators, $\tau_1$, $\tau_2$
and $\tau_3$.  One can easily check that with the choice of the ground
state   
(\ref{choice}) the expectation value of $[\tau_1,\tau_2]=i\tau_3$ is
non-vanishing so that the usual counting of Goldstone modes cannot be
used.\footnote{The ground state has nonzero isospin density because
the condensed kaon has a definite isospin.} (Our theorem however
does not {\em require} the number of Goldstone modes to be different 
from three).   

This phenomenon was not observed in earlier studies of meson
condensation in QCD \cite{KSTVZ,Misha-Son,Kim}. For example, in the
case of a chemical potential for isospin in QCD with $N_c=3$ and
$N_f=2$ meson condensation only leads to the breakdown of a global
$U(1)$, so there is only one broken generator and one Goldstone
mode. Also the case of QCD with $N_c=2$ colors and $N_f=2$ flavors at
finite baryon density  
we have only one Goldstone boson due to the spontaneous breaking of
the $U(1)$ baryon symmetry by the diquark condensate.

In general we can write the determinant of the inverse propagator as
\be \det D = \prod_k (p^2 -E^2 -\lambda_k). \ee  
The dispersion relations are obtained from $\det D =0$.
For a charge
conjugation invariant system, $\det D$ is an even function of $\mu$,  
and, for finite quark masses, a smooth function
of $\mu$ in the neighborhood of $\mu =0$. 
Since $\mu$ only occurs in the combination $\mu E$ in
the coupling matrix elements, we conclude that $\det(D)$ is a
polynomial in $E^2$.  Suppose that we have a curvature matrix with   
$n$ zero eigenvalues and $n_1$ ($n_2$) of these eigenvalues vanish linearly 
(quadratically) for small $E$. 
For each
eigenvalue that approaches zero  linearly in $E$ 
we find one mode with a quadratic 
dispersion relation and one mode with a nonzero mass, and, for each
eigenvalue that approaches zero quadratically in $E$,  
we find two modes with a linear
dispersion relation. Using charge conjugation invariance properties of our theory we find that the  
Nielsen-Chadha theorem \cite{holger-counts} in this case reduces to
the simple statement that $n_1+n_2 = n$.

\section{Kaon Condensation in QCD}

Let us finally consider the low-energy limit of QCD for
$N_f$ light quarks and a nonzero
chemical potential $\mu$ for one of these quarks. For convenience, this 
quark will be identified as a strange quark. 
It is now well understood how to extend a chiral Lagrangian to nonzero
chemical  potential \cite{kogut1,KSTVZ,DT-phase,Misha-Son,KT,Kim,STV}:
Requiring QCD and the low-energy effective Lagrangian
to have the same transformation properties of a local external vector source 
uniquely determines the dependence
of the chiral Lagrangian on this vector source and thus on the
chemical potential.  No additional parameters are necessary. The
low-energy  limit of QCD with a nonzero strangeness chemical potential
is thus given   by the Lagrangian, 
\be 
{\cal L} = \frac {F^2}4 {\rm Tr}
 \nabla_\nu \Sigma \nabla_\nu \Sigma^\dagger
-\frac 12 G {\rm Tr} ({\cal M}^\dagger \Sigma + {\cal M}\Sigma^\dagger),
\ee
where the $SU(N_f)$ valued field $\Sigma$ contains the
pseudo-Goldstone modes and ${\cal M}$ is the mass matrix which is
taken proportional to the identity in this note.
This theory contains two parameters,
the pion decay constant $F$ and  the chiral condensate $G$.
The chemical potential enters through the covariant
derivative
\be
\nabla_\nu\Sigma = \partial_\nu \Sigma -i [B_\nu, \Sigma],
\ee
where $B_\nu$ only contains the chemical potential for the 
strange quark  and is given by 
\be 
 B_\nu = \delta_{0\nu} {\rm diag}(0,\cdots,0,-i\mu). 
\ee
Kaon-condensation in QCD at very high density is described by the same
chiral Lagrangian with the chemical potential replaced by an induced
chemical potential $\mu_{\rm eff}\simeq m_s^2/2p_F$. The covariant 
derivative of the chiral field takes the form 
\be
 \nabla_\nu \Sigma = \partial_\nu \Sigma
   + \delta_{\nu 0} \left( \frac{{\cal MM}^\dagger}{2p_F}\Sigma
        - \Sigma^\dagger \frac{{\cal M^\dagger M}}{2p_F} \right). 
\ee
In the CFL phase, the structure of the mass term is also
different. Because of color-flavor-locking, we have to  
replace ${\cal  M} \to \tilde{\cal M}=\det({\cal M}){\cal M}^{-1}$ 
\cite{Misha-Son-inverse}.

In \cite{KT}  this theory was studied in detail for $N_f = 3$. It was
found that a  phase transition to a kaon condensed phase takes place
at  $\mu = M$. For $\mu < M$ the saddle point of the static
part  of the effective Lagrangian is given by the identity matrix.  
The pole masses of the pseudo-Goldstone modes are modified by the
chemical  potential according to  
\be M \rightarrow M- S \mu,  \qquad \mu <
M, \ee  
where $S$ is the strangeness of the Goldstone bosons. These
pseudo-Goldstone modes satisfy standard dispersion relations with
$E^2 -(\sqrt{p^2+M^2} -S\mu)^2$=0.  
For $N_f =3$ two pseudo-Goldstone bosons become massless at
$\mu=M$. For general value of $N_f$, $N_f -1$ Goldstone bosons become
massless at the critical point.   

In the kaon condensed phase both the quark-antiquark condensate and 
the kaon condensate are nonzero. The saddle point manifold has 
the structure \cite{DT-phase,Misha-Son,KT} \be \bar \Sigma = 
\left ( \begin{array}{ccccc} 1  & \cdots& 0&0& 0 \\
                                        \vdots & \ddots& \vdots& \vdots 
                                         &\vdots \\
                                        0& \cdots &1&  0& 0  \\
                           0 & \cdots &0& \cos \alpha & e^{i\theta} \sin \alpha\\
                           0 & \cdots & 0&- e^{-i\theta}
                               \sin \alpha & \cos \alpha \\
       \end{array}\right ),
\ee
where $\cos \alpha=M^2/\mu^2$ for $\mu>M$.
The iso-vector symmetry of the QCD Lagrangian with non-zero quark mass
and non-zero strange chemical potential is thus broken spontaneously
according to   
\be
SU(N_f-1)\times U(1) \to SU(N_f -2) \times U(1).
\ee
According to the naive version of 
Goldstone's theorem the total number of massless states
is equal to $(N_f-1)^2 -(N_f-2)^2 = 2N_f-3$. We thus 
find a mismatch in the total
number of massless states on either sides of the phase transition point. 
In the same way as was
discussed before this mismatch is resolved because of the presence of
Goldstone  modes with a quadratic dispersion relation.

However, this leads to another puzzle. 
Our model describes a continuous phase transition with a saddle point
that is a continuous function of the chemical potential. The
eigenvalues of the  second derivative matrix at the saddle point and  
the corresponding poles masses should be continuous at the phase
transition  point as well. 
However, while above the critical point there is a Goldstone mode with
linear dispersion, right at the transition point all gapless degrees
of freedom have quadratic dispersion.  Indeed, at that point 
the two poles with non-zero strangeness quantum number
obey the dispersion relation
\be
E^2 = [ \sqrt{ M^2+p^2}\pm M]^2 = 2M^2 +p^2 \pm 2 M\sqrt{M^2+p^2}. \ee
The dispersion relation of the would-be Goldstone modes  is thus given
by \be E^2 = \frac {p^4}{4M^2}+ \cdots  \, ,
\ee
which is quadratic.

The solution to this puzzle is simple:
continuity in $\mu$ requires that a linear 
dispersion relation for $\mu > \mu_c$
is given by
\be
E^2 = \gamma(\mu) p^2 +  \frac {p^4}{4M^2} + \cdots \qquad  
{\rm for}\quad \mu> \mu_c, \ee 
where $\gamma(\mu_c)=0$. Indeed, this general result is in agreement 
with the dispersion relations (\ref{displ},\ref{dispq}) of our 
model Lagrangian (\ref{Lagr}). 

Above the  phase transition point in the superfluid phase, the matrix of second  
derivatives factorizes into two $4 \times 4$ matrices given by
\cite{KT} \be M_1 =  \frac 12 \left ( \begin{array}{cccc} E^2 -p^2 
    -\mu^2 & 0  & 0 & 0 \\
       0 & E^2 -p^2 -\frac{2 M^2+\mu^2}{3 \mu^2}& 
                         -\frac 2{\sqrt 3} \mu \sin\alpha & 0 \\
                               0 & -\frac 2{\sqrt 3} \mu \sin\alpha 
& E^2 -p^2 &  
                     -2iE\mu\cos\alpha \\
                                           0 & 0 & 2iE\mu\cos\alpha & E^2 - 
p^2-\mu^2\sin^2\alpha \\
                       \end{array} \right ),\nonumber \\
\ee
and 
\be
M_2 = \frac 12 \left ( \begin{array}{cccc} E^2 -p^2 -M^2 & 
i E\mu(\cos \alpha -1) &  E\mu \sin \alpha & 0 \\
            -i E\mu(\cos \alpha -1)     & E^2 -p^2 -M^2 & 0 & 
               - E\mu \sin \alpha \\ 
 E\mu \sin \alpha  & 0 & E^2 -p^2 
& i E \mu (\cos\alpha+1) \\0 &
       -E \mu \sin \alpha & 
     -i E \mu (\cos\alpha+1)  & E^2 - p^2 \\
                       \end{array} \right ).\nonumber \\
\ee
At the transition point the
two matrices are the same and the determinant agrees with the
determinant  obtained from the determinant of the would be zero modes
in the  normal phase. The dispersion relation of the modes is obtained
by  calculating the zeroes of the determinant of $M_1$ and $M_2$ which
can be  
obtained  perturbatively for  $\mu$ approaching $\mu_c$. The
dispersion  relations obtained from the determinant of $M_1$ are given
by   
\be E^2 &=& \bar \mu p^2 ,\\ E^2 &=& M^2 + p^2+2 \bar \mu M^2,\\ E^2
&=& M^2 + p^2  +\bar \mu(-\frac {22}9 M^2 + \frac{16}{27} p^2),\\ E^2
&=& 4 M^2+2p^2 
+\bar \mu (\frac {28}9 M^2 -\frac{43}{27} p^2), \ee
where $\bar \mu = 
 (\mu - \mu_c)/M$. For $M_2$ we find the dispersion relations 
\be E^2
 &=&  \frac 14 M^2 (1-\bar \mu^2) p^4,\\ E^2 &=& M^2 +p^2 +\bar  \mu(-2M^2 + p^2), \\  
E^2 &=& M^2+p^2 +\bar \mu(-\frac 23 M^2 -\frac 19 p^2), \\ 
E^2 &=& 4M^2+ 2p^2 +\bar \mu (\frac {32}3 M^2 - \frac 89 p^2). \ee In
both  cases the modes match up to Goldstone modes in the normal phase
with  strangeness equal to $+1$, 0, 0 and $-1$, respectively. At the
transition  point we observe one Goldstone boson with a linear
dispersion  relation and one with a quadratic dispersion
relation. This requires  that the total number of broken generators is
equal to  three which is indeed the case. In agreement with the  
above continuity argument, the coefficient of the linear dispersion
relation vanishes at the critical point.

As a side remark we next answer the question of what is the 
simplest effective Lagrangian that described the massless states of 
the theory.
General arguments show that an effective Lagrangian 
for a particle
with a quadratic dispersion relation and a kinetic term that is
quadratic in  the momenta requires at least two degrees of
freedom. The  simplest effective Lagrangian describing our massless
state  thus has to contain at least three degrees of freedom. We thus
 conclude that the simplest effective Lagrangian for the massless
 modes  at the transition point is given by 
\be {\cal L} = \frac 12
 \partial_\nu  \Sigma \partial_\nu \Sigma^\dagger 
+\mu{\rm Tr} \Sigma^\dagger V\partial_0 \Sigma,
\ee
where $\Sigma \in SU(2)$ and $V={\rm diag}(0,1)$.  This Lagrangian
might be of interest for phenomenological applications of the results
obtained in this paper. 

\section{Conclusion}
In conclusion, in the kaon condensed phase we have found one Goldstone
mode with a linear dispersion relation and all others with a quadratic
dispersion  relation. A mismatch in the naive counting of the
Goldstone modes  is resolved by the observation that in the latter
case the  number of Goldstone modes is only half the number of broken
generators.

\vskip 0.5cm
\noindent{\bf Acknowledgements}:
P. Bedaque, E. Fradkin, N. Goldenfeld, J. Goldstone, D.-K. Hong,
P.A. Lee and W.V. Liu  are thanked for useful discussions.   
This work was partially supported by the US DOE grant DE-FG-88ER40388.
T.S., D.T.S., and M.A.S. thank RIKEN, Brookhaven National Laboratory,
and U.S.\ Department of Energy [DE-AC02-98CH10886] for providing the
facilities essential for the completion of this work. D.T.S. is
supported, in part, by an DOE OJI grant.  
D.T. is  supported, in part, by
``Holderbank''-Stiftung.  

After this work was completed we became 
aware of Ref.\ \cite{miransky} in which the dispersion relations of
the model (\ref{Lagr}) were derived as well.

\end{document}